\documentclass[final]{ias2}

\usepackage{graphicx} 
\usepackage{multirow}
\usepackage{array} 

\usepackage{hyperref} 
\newcommand{\pp}{$pp$}
\newcommand{\pA}{$pA$}
\newcommand{\dA}{$dA$}
\newcommand{\pdA}{$p(d)A$}
\newcommand{\Aa}{{\rm AA}}
\newcommand{\ppb}{$p$+Pb}
\newcommand{\dau}{$d$+Au}
\newcommand{\pbpb}{Pb+Pb}
\newcommand{\auau}{Au+Au}
\newcommand{\gev}{GeV/$c$}
\newcommand\pt{p_T}
\newcommand\ptt{p_T^{\rm (t)}}
\newcommand\pta{p_T^{\rm (a)}}
\newcommand\dphi{\Delta\phi}
\newcommand\deta{\Delta\eta}
\newcommand\kt{$k_T$}
\newcommand\mean[1]{\langle#1\rangle}

\begin{document}

\markboth{Jet Quenching and Correlations}{Fuqiang Wang}

\title{Jet Quenching and Correlations}

\author[purdue]{Fuqiang Wang} 
\email{fqwang@purdue.edu}
\address[purdue]{Department of Physics, Purdue University, West Lafayette, Indiana 47907, USA}

\begin{abstract}
This article reviews recent advances in our understanding of the experimental aspects of jet-quenching and correlations in relativistic heavy-ion collisions at RHIC and LHC. Emphasis is put on correlation measurements, namely jet-like correlations with anisotropic flow subtraction in heavy-ion collisions and long-range pseudorapidity correlations in small systems. Future path on correlation studies is envisioned which may elucidate jet-medium interactions and the properties of the hot dense medium in QCD. 
\end{abstract}


%\pacs{25.75.-q, 25.75.Bh, 25.75.Gz, 25.75.Ld}
 
\maketitle

% \tableofcontents
% \listoffigures
% \listoftables

\section{Introduction}

The primary goal of relativistic heavy-ion collisions is to create a state of matter at high energy density and temperature where quarks and gluons are deconfined over an extended volume of the order of the nuclear size. Such a state is called the quark-gluon plasma (QGP). Experimental evidence indicates that a QGP is created at BNL's Relativistic Heavy-Ion Collider (RHIC)~\cite{Arsene:2004fa,Back:2004je,Adams:2005dq,Adcox:2004mh} and at CERN's Large Hadron Collider (LHC)~\cite{Muller:2012zq}. The created QGP is strongly interacting and behaves like a nearly perfect fluid~\cite{Gyulassy:2004zy,Tannenbaum:2006ch}. The created QGP provides a test ground to study quantum chromodynamics (QCD)--the fundamental theory governing the strong interaction among quarks and gluons--under extreme conditions of high energy density and temperature~\cite{Jacobs:2004qv}.

Two major experimental observations are essential in establishing the perfect fluid paradigm of the strongly interacting QGP~\cite{Gyulassy:2004zy,Tannenbaum:2006ch}. One is the large collective (radial and anisotropic) flow measured by final-state low transverse momentum ($\pt$) particles~\cite{Heinz:2013th}. The collective flow is generated by hydrodynamic expansion powered by the high pressure buildup in the central region of the heavy-ion collision zone~\cite{Ollitrault:1992bk,Huovinen:2006jp}. The other is the strong suppression of high-$\pt$ particle yields and correlations in the final state~\cite{Jacobs:2004qv}. The suppression arises from attenuation of energetic partons losing energy due to interactions with the QGP medium, via gluon bremsstrahlung and multiple scattering~\cite{Gyulassy:1990ye,Wang:1991xy,Gyulassy:2003mc}. The initial energetic partons are produced by hard-scatterings and fragment into jets of collimated particles in the final state. The high-$\pt$ suppression phenomenon is thus often called jet-quenching~\cite{Gyulassy:1990ye,Wang:1991xy}, and can be studied in great detail by particle angular correlations~\cite{Wang:2013qca}.

This article reviews experimental observations of jet-quenching and correlations, with an emphasis on two-particle angular correlations at low to intermediate $\pt$. The review is organized as follows. Section~\ref{sec:theory} gives a brief, phenomenological description of the theory behind jet-quenching from an experimentalist perspective. Section~\ref{sec:highpt} reviews high-$\pt$ suppression measurements, highlighting the recent measurements of jets at the LHC, followed by Sec.~\ref{sec:gamma} on jet-hadron, $\gamma$-hadron and $\gamma$-jet correlations. Section~\ref{sec:dihadron} discusses dihadron correlations at low-intermediate $\pt$ in heavy-ion collisions, focusing on the interplay between anisotropic flow and jet-like correlations, followed by Sec.~\ref{sec:ridge} on the long-range pseudorapidity (ridge) correlations in small systems. %Section~\ref{sec:future} tries to point out broad directions of future measurements of particle correlations. A summary is given in Sec.~\ref{sec:conclusion}.
Section~\ref{sec:future} summarizes the article with future prospects of jet-quenching and particle correlation measurements.

\section{Jet-quenching and jet modification\label{sec:theory}}

QCD governs the interactions of quarks and gluons. The QCD Lagrangian is well defined~\cite{Dokshitzer:1991wu,Qiu:2011zza}. Because gluons are self-interacting, the coupling constant at typical momentum transfers of parton-parton interactions is large, $\mathcal{O}(1)$. At this large coupling, QCD is analytically incalculable and one has to resort to computer simulations on a lattice~\cite{Fodor:2009ax}. At large momentum transfer, the coupling constant is small (asymptotic freedom) and perturbative theory can be exploited to calculate QCD processes (pQCD)~\cite{Dokshitzer:1991wu,Qiu:2011zza}. Large momentum transfer in a parton-parton scattering results in energetic partons at large angles with respect to the interaction (beam) axis. These energetic partons fragment ultimately into jets of hadrons--color singlets that can exist in vacuum. In terms of kinematics, a jet in the final state and its colored parton originator are almost synonymous, and thus the words of ``jet'' and ``parton'' are often used interchangeably. 

The production cross-section of final-state particles from hard-scattering processes can be schematically represented in the form:
\begin{eqnarray}
\frac{d\sigma_h}{d\pt}=\sum_{abc}\int{dx_adx_b} & f_A(x_a,Q^2)f_B(x_b,Q^2)\times\hat{\sigma}_{ab\to c}(x_a,x_b;Q^2,\alpha_s)\times\nonumber\\
&D_{c\to h}(z_{h/c},Q^2)\,.
\label{eq:qcd} 
\end{eqnarray}
It is a convolution of a short-distance perturbative cross-section, $\hat{\sigma}$, nonperturbative parton distribution functions, $f(x,Q^2)$, and jet fragmentation function, $D(z_{h/c},Q^2)$. The sum runs over all parton species involved in scattering processes, $a+b\to c$, that can produce a final-state hadron, $h$, from the fragmentation of parton $c$. $A$ and $B$ denote the colliding projectile and target, and $x_a$ and $x_b$ are the momentum fractions carried by the incident partons. The variable $z_{h/c}$ is the longitudinal fraction of parton $c$ momentum carried by the hadron $h$; it contains the final hadron $\pt$ information. $Q$ is the momentum transfer and $\alpha_s$ is the strong coupling constant. The cross-section can be calculated by pQCD~\cite{Dokshitzer:1991wu,Qiu:2011zza} at small $\alpha_s$ (at large $Q$). The structure function and the fragmentation function are non-perturbative and cannot be calculated by pQCD, but each can be evolved from a starting distribution at a defined energy scale~\cite{Dokshitzer:1991wu,Qiu:2011zza}.
%\note{Perturbative QCD calculations may have colored partons in the final state, but only the colorless hadrons they ultimately produce are observed experimentally. Thus, to describe what is observed in a detector as a result of a given process, all outgoing colored partons must first undergo parton showering and then combination of the produced partons into hadrons. The terms fragmentation and hadronization are often used interchangeably in the literature to describe soft QCD radiation, formation of hadrons, or both processes together.}
%\note{As the parton which was produced in a hard scatter exits the interaction, the strong coupling constant will increase with its separation. This increases the probability for QCD radiation, which is predominantly shallow-angled with respect to the originating parton. Thus, one parton will radiate gluons, which will in turn radiate qq pairs and so on, with each new parton nearly collinear with its parent. }

The underlying hard-scattering cross-section is the same in proton-proton (\pp) and heavy-ion (\Aa) collisions. The initial-state parton distributions are not expected to be vastly different between proton and nucleus (see below). In the final state, a QGP medium is not expected to form in minimum-bias \pp\ collisions. Whereas in \Aa\ collisions, a QGP is formed, on the time scale of 1~fm/$c$ with a size of several fm (nucleus size). Because short-distance hard-scatterings happen early in time during a collision, the hard-scattering partons have to traverse the hot and dense QGP medium and are expected to interact with the medium and lose energy~\cite{Gyulassy:1990ye,Wang:1991xy}. The fragmentation of the reduced-energy parton will yield fewer particles at high $\pt$ in the final state. 
Differences between \pp\ and \Aa, therefore, primarily result from a change in the final state, or the fragmentation function $D_{c\to h}(z_{h/c},Q^2)$ in Eq.~(\ref{eq:qcd}). A comparison of the final-state high-$\pt$ particle yields in \pp\ and \Aa\ collisions will thus reveal the effect of jet-quenching. 
Two descriptions are often used:
%Often, two pictures/languages are used to describe the same phenomenon: 
(i) Partons are produced identically in \pp\ and \Aa; partons fragment in vacuum in the \pp\ case, and in the \Aa\ case partons fragment in the presence of the medium differently than in vacuum. This is often called modifications to jet fragmentation by the heavy-ion medium. (ii) Partons produced in \Aa\ collisions lose energy via gluon radiation and multiple scattering in the medium; they then exit the medium with reduced energy and fragment in vacuum as in \pp.
These two descriptions, however, refer to the same phenomenon of partonic energy loss.

\section{High-$\pt$ suppression measurements\label{sec:highpt}}

As discussed in Sec.~\ref{sec:theory}, jet-quenching manifests itself as a suppression of final-state high-$\pt$ particle yields. 
High-$\pt$ particle invariant yields were measured at RHIC~\cite{Adcox:2001jp,Adler:2002xw,Adler:2003qi,Adams:2003kv} and the LHC~\cite{Aamodt:2010jd,CMS:2012aa}. Strong suppressions were observed in central heavy-ion collisions relative to \pp\ collisions. This indicates that jet fragmentation is softened in central heavy-ion collisions. It is, however, still possible that differences in the initial-state parton distributions in proton and nucleus, expected in QCD, could play an important role. %This is because, although the parton distribution functions are not expected to be vastly different between proton and nucleus, difference exists based on QCD and could yield significant different in the final-state hadron distribution. 
The question has to be settled experimentally. For this reason \dau~\cite{Adler:2003ii,Adams:2003im} and \ppb~\cite{ALICE:2012mj} collisions were conducted and it was found that high-$\pt$ particles were not suppressed in \dau\ relative to \pp\ collisions. This demonstrates that the large suppression observed in central heavy-ion collisions must arise from final-state interactions--jet-quenching due to partonic energy loss. Referring to Eq.~(\ref{eq:qcd}), the parton fragmentation functions are different between the \pp\ and \Aa\ cases. 

Single particle measurements give a convoluted information. A more direct measurement of jet fragmentation function, and its modification in heavy-ion collisions, is through fully reconstructed jets~\cite{Cacciari:2010te}. Because jet production cross-section is given only by the $\hat\sigma$ and $f$ terms in Eq.~(\ref{eq:qcd}), without the $D$ term, cross-section measurements of fully reconstructed jets in \pp\ and \Aa\ should be equal (except the small initial-state differences). Ideally, fully reconstructed jets should yield a complete picture of the effect of jet-quenching and how the lost energy is redistributed. In reality, however, jet reconstruction carries with it a set of shortcomings. For example, to reduce background and background fluctuations, often a high-$\pt$ particle is required as a seed in jet reconstruction, biasing jets toward large energies and surface emission; often a small jet cone size has to be used, resulting in incomplete jets. An extreme would be jet composed of a single high-$\pt$ particle. As a result, low-energy jets and low-$\pt$ jet fragments cannot be reliably reconstructed. The reconstructed jets are not full jets, but partial and biased. This limits the power of reconstructed jets to study jet-quenching because the very effect of jet-quenching is the transport of energy to lower $\pt$ and larger angles. On the other hand, these biases can be used to study jet-quenching, for example, by measuring cross-section suppression of reconstructed jets~\cite{Aad:2012vca}, by studying the energy imbalance of reconstructed back-to-back jets~\cite{Aad:2010bu,Chatrchyan:2011sx,Chatrchyan:2012nia}, and by measuring jet cross-section as a function of the cone radius parameter used in jet reconstruction~\cite{Putschke:2008wn,Ploskon:2009zd,Aad:2012vca,Reed:2013rpa}.

Calorimeter jets were reconstructed by ATLAS~\cite{Aad:2012vca} using the anti-\kt\ algorithm~\cite{Cacciari:2008gp} with a cone radius parameter $R=\sqrt{\dphi^2+\deta^2}$ varying from 0.2 to 0.5. A suppression of a factor of 2 in jet cross-section is observed in top 10\% central \pbpb\ collisions relative to 60-80\% peripheral collisions. The suppression is observed to be insensitive to the reconstructed jet energy, to increase with the collision centrality, and to decrease with increasing $R$. This is a direct evidence of jet quenching where fewer jets are reconstructed at a given energy within the limited $R$ range.

The momentum asymmetry between a leading jet and its back-to-back partner jet was measured~\cite{Aad:2010bu,Chatrchyan:2011sx,Chatrchyan:2012nia}. It is found that the asymmetry is larger in central \pbpb\ collisions than in \pp\ collisions. This is consistent with jet-quenching: the subleading partner jets, being less surface biased, have interacted with the medium more than the leading jets; this causes more energy to transport to outside the limited cone size, and/or more particles falling below the $\pt$ cut-off used in jet reconstruction.

Reconstructed jets give direct access to jet fragmentation function--the energy partition of the constituents used in the reconstructed jet. 
%
%The measurement of single particle production offers limited information. In order to gain further insights into modifications of jet fragmentation and jet shape, 
%To this end, full jet reconstruction~\cite{Cacciari:2008gp,Cacciari:2010te} is carried out in experiments~\cite{otherExp,}. 
%In order to overcome the large background in heavy-ion collisions, a lower $\pt$ cut-off is often imposed on particles to be collected as constituents of jets. Also because of the large heavy-ion background, jets reconstructed at low energy is severely contaminated by fake jets and thus often only high energy jets are reported. 
%
The jet fragmentation functions have been measured by CMS~\cite{Chatrchyan:2012gw}. % are shown in Fig.~\ref{fig:CMSjetfrag}. 
Jets were reconstructed from tracking and calorimetric information using the anti-\kt\ algorithm~\cite{Cacciari:2008gp} with a cone radius parameter of 0.3. To minimize background from the underlying events, only tracks with $\pt>4$~\gev\ were used in jet fragmentation measurements. The jet fragmentation functions in \pbpb\ collisions are observed to be consistent with that in \pp\ collisions. This indicates that the lost energy by energetic partons must reside at angles larger than 0.3 and/or in $\pt$ lower than 4~\gev.
%a modest modification of jet fragmentation in central \pbpb\ collisions, with energy moved from intermediate $\pt$ fragments to lower $\pt$ fragments. 
%Measurements of jets at the LHC indicate that a depletion of jet fragments in the intermediate $z$ and enhancement of low $z$ (low $\pt$ for mid-rapidity jets) particles. \note{why z=1 not suppressed, trigger bias?} 

%\begin{figure*}
%\begin{center}
%\includegraphics[width=0.9\textwidth]{CMS_jetfrag.pdf}
%\caption{(a,b) Jet fragmentation functions reconstructed in peripheral and central \pbpb\ data for the leading (open circles) and subleading (solid points) jets. (c,d) Ratio of each \pbpb\ fragmentation function to its \pp-based reference. Error bars are statistical. The hollow boxes in panels c and d represent the systematic uncertainty for the leading jet, and gray boxes show the systematic uncertainty for the subleading jet. From CMS~\cite{Chatrchyan:2012gw}.}
%\label{fig:CMSjetfrag}
%\end{center}
%\end{figure*} 

\section{Jet-hadron, $\gamma$-hadron, and $\gamma$-jet correlations\label{sec:gamma}}

%Dihadron correlations with identified particles can provide further information of partonic energy loss mechanisms. Identified particle correlations have been measured by several experiments~\cite{Adler:2004zd,}.

\begin{figure*}
\begin{center}
\includegraphics[width=1.03\textwidth]{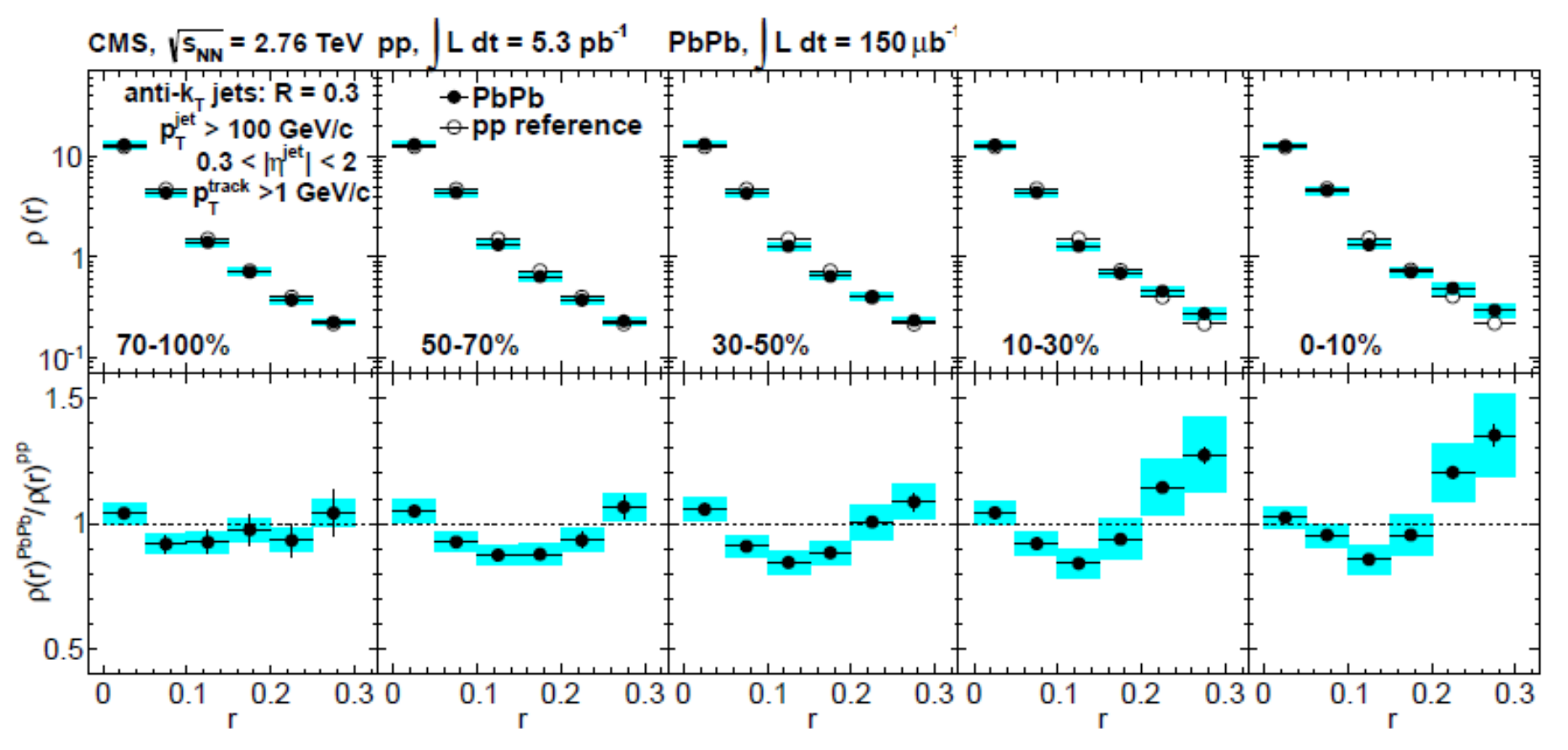}
\caption{(Upper panels) Differential jet shapes in \pbpb\ collisions (filled circles) as a function of angular distance from the jet axis for inclusive jets with $\pt>100$~\gev\ and $0.3<|\eta|<2$ in five \pbpb\ centrality intervals. The measurements use charged particles with $\pt>1$~\gev. The \pp-based reference shapes are shown with open symbols. The shaded regions represent the systematic uncertainties for the measurement performed in \pbpb\ collisions, with the statistical uncertainties too small to be visible. (Lower panels) The ratio of jet shape in \pbpb\ to that in \pp. The error bars show the statistical uncertainties, and the shaded boxes indicate the systematic uncertainties. From CMS~\cite{Chatrchyan:2012gw}.}
\label{fig:CMSjetshape}
\end{center}
\end{figure*}

In order to investigate where the lost energy goes, one may carry out jet-hadron correlation analysis. In such analysis, low-$\pt$ particles are correlated with reconstructed jets, statistically, even though they are not assigned to jets event-by-event. CMS measured jet-hadron correlations as a function angular distance from the jet axis~\cite{Chatrchyan:2011sx}. %The underlying event background is subtracted using particles at the opposite $\eta$ about $\eta=0$ but at the same $\phi$. 
To reduce the background, only tracks with $\pt>1$~\gev\ were used in the CMS analysis. It was observed that the correlated particles at low $\pt$ appear at relatively large angle in \pbpb\ collisions, especially for subleading jets~\cite{Chatrchyan:2011sx}. At high $\pt$ no noticeable changes are observed, consistent with the fragmentation measurements~\cite{Chatrchyan:2012gw}. 
Jet shape measurements were also carried out for inclusive jets by CMS~\cite{Chatrchyan:2013kwa}.
%A similar jet shape analysis was carried out by CMS where the angular density of fragment $\pt$ with jets is measured as a function of the angular distance from the jet axis (jet shape) (Ejet, track pt cut, pttrack>1~\gev). 
Figure~\ref{fig:CMSjetshape} shows the jet shapes in \pp\ and for five centralities in \pbpb\ collisions in the upper panels, and the ratio of jet shapes in \pbpb\ to \pp\ in the lower panels.
Modifications are observed in \pbpb\ collisions where jet energy is moved from small angles to large angles. 
%The CMS measurements suggest that the lost energy in the very soft mode appears to be permanently lost, over the entire $2\pi$ azimuth, indistinguishable from the medium.

%To check momentum balance of dijets including particles of all $\pt$, the CMS experiment investigated momentum balance of dijets by...  Although the jets were reconstructed with a relatively large $\pt$ cut and small cone size, all~\cite{Chatrchyan:2011sx}. By subtracting subtracting the away-side....

%Reconstructed jets can be used as the trigger in jet-hadron correlation studies. 
Jet-hadron correlation results from STAR~\cite{Adamczyk:2013jei} indicate that energy is shifted from high-$\pt$ to low-$\pt$ fragments and the jet shape is broadened. The systematic uncertainties are presently large due to uncertainties in elliptic and triangular flows of jets in background subtraction. 
%Jet-hadron corrrelations offer a more direct information of the trigger parton kinematics and direction. However, what have been learned so far from jet-hadron correlations are limited, compared to high-$\pt$ trigger dihadron correlations, given the systematics involved in full jet reconstruction.

Direct $\gamma$-jet and $\gamma$-hadron correlations are still the golden probe of jet-quenching~\cite{Wang:1996yh,Zhang:2009rn,Li:2010ts}. Direct photons, once produced, do not interact with the medium via the strong interaction. Their energy is unaltered providing a gauge of the away-side jet energy. Another advantage is that direct photons are azimuthally isotropic, immune from complications of flow background subtraction. Direct photons are, however, notoriously difficult to measure because of their rare production and large contaminations from decays, predominantly of $\pi^0$'s. Separating direct photons from other sources, such as fragmentation photons and photons produced by jet-medium interactions (bremsstrahlung and conversion)~\cite{Qin:2009bk}, is also challenging.

Direct $\gamma$-hadron correlations have been measured at RHIC~\cite{Adare:2009vd,Abelev:2009gu,Adare:2012qi}. Suppression of high-$\pt$ associated particles apposite to a $\gamma$ trigger is observed to be similar to those opposite to a hadron trigger. PHENIX measurement~\cite{Adare:2012qi} at low associated $\pt$ indicates that the low-$\pt$ associated yield is enhanced and the enhancement is dominated by large angles. These observations are also consistent with those by dihadron correlations. 

The similarities between $\gamma$-hadron and dihadron correlations are not expected a priori. There are expected physics differences between $\gamma$-hadron and dihadron correlations. (i) Direct photons are primarily produced by gluon compton scattering, $g+q\to\gamma+q$, at RHIC energy in the central rapidity region. The recoil jet opposite to a direct $\gamma$ is thus dominated by quarks, whereas dihadron correlations are dominated by gluons. Energy losses by quarks and gluons are expected to be different. (ii) A leading hadron is a fragmentation product of a jet, carrying only a fraction of the parent parton, whereas direct photons carry the entire energy. (iii) High-$\pt$ trigger particles in dihadron correlation measurements are biased toward surface emission due to energy loss, maximizing the path-length the away-side jet has to traverse. Since photons do not interact with the medium via the strong interaction, photons are emitted over the entire volume of the collision zone. They may be even biased toward the opposite surface at relatively large associated particle $\pt$, because those associated particles are surface biased in their emission.

Even though the $\gamma$-hadron correlation data are similar to dihadron data, the statistical and systematic uncertainties of the $\gamma$-hadron correlations are presently large. This calls for high statistics $\gamma$-hadron correlation measurements in the future.

Direct $\gamma$-jet correlations were measured at the LHC in \pp\ and \pbpb\ collisions at midrapidity~\cite{Chatrchyan:2012gt}. The energies of the direct $\gamma$ and jets (anti-\kt, cone radius parameter $R=0.3$) are required to be at least 60~\gev\ and 30~\gev, respectively. Significantly degrading of the partner jet energy was observed in central \pbpb\ collisions, consistent with jet-quenching. However, no angular broadening was observed in the $\gamma$-jet correlation functions.

\section{Dihadron correlations in heavy-ion collisions\label{sec:dihadron}}

Jet reconstruction yields event-by-event information of jets but has its own drawbacks. The average properties of jets, however, can be obtained from dihadron correlations with high-$\pt$ trigger particles. Because of the requirement of high-$\pt$ trigger particles, the jets probed by dihadron correlations are a biased sample of jets which happen to have a hard fragmentation--the trigger particle typically takes away $\mean{z}=70\%$ of the jet energy~\cite{Boca:1990rh,Kopeliovich:2012sc}--and are preferentially emitted from the surface region directed outward~\cite{Jacobs:2004qv}. This is similar to the jet reconstruction case where a high-$\pt$ particle is required as an initial seed. Without a high-$\pt$ trigger particle, dihadron correlations are dominated by non-jet physics, although jets do fragment into low-$\pt$ soft particles~\cite{Agakishiev:2011pe}. 

A two-particle correlation function is typically defined as the two-particle density normalized by the product of two single particle densities. Experimentally, two-particle correlation is constructed by the ratio of two-particle density in real event to that in mixed events (which are constructed from different events of the same characteristics). Namely,
\begin{equation}
\frac{d^2N(\deta,\dphi)}{d\deta d\dphi}=\frac{d^2N_{\rm real}}{d\deta d\dphi}\cdot\frac{\left.d^2N_{\rm mixed}/d\deta d\dphi\right|_{\rm max}}{d^2N_{\rm mixed}/d\deta d\dphi}\,.
\label{eq:corr}
\end{equation}
In analysis of two-particle correlations with trigger particles, the purpose is to study particle yield correlated to the trigger particle. Thus the mixed-event two-particle density in Eq.~(\ref{eq:corr}) is treated as acceptance correction after proper normalization to unity at maximum acceptance, e.g.~at $(\deta,\dphi)=(0,0)$. Note that the proper acceptance correction should be the convolution of two single-particle detector acceptance$\times$efficiencies. For approximately uniform $dN/d\eta$ or integrated trigger and associated particle $\eta$ ranges, the acceptance correction by mixed-events is a good approximation~\cite{Xu:2013sua}.

Most of the particles produced in a heavy-ion collision are unrelated to the trigger particle, but combinatorial background. The two-particle correlation function can be written into two parts, background and signal:
\begin{eqnarray}
\frac{d^2N(\deta,\dphi)}{d\deta d\dphi}&=&B_2\left(1+2\sum_{n=1}^{\infty}V_n(\deta;\ptt,\pta)\cos n\dphi\right)+\nonumber\\
&&\frac{d^2N_{\rm signal}(\deta,\dphi)}{d\deta d\dphi}\,,
\label{eq:2p}
\end{eqnarray}
where $B_2$ is the average background pair density $\mean{d^2N_{\rm bkgd}/d\deta d\dphi}$. Note that the $\pt$ variables of the trigger and associated particles are suppressed in Eqs.~(\ref{eq:corr}) and (\ref{eq:2p}) except in the Fourier coefficient $V_n$. The background level is not known a priori. An {\em ad hoc} procedure is often used, called ZYAM~\cite{Adams:2005ph,Ajitanand:2005jj}, which assumes the jet-correlation signal has zero yield at minimum.
In order to extract jet-correlated information, $d^2N_{\rm signal}/d\deta d\dphi$, the combinatorial background needs to be subtracted. Because the background is nonuniform in $\dphi$ and the modulation is not precisely known, the background subtraction has a large uncertainty when the overall background level is large. To reduce the background, one may go to high associated $\pt$~\cite{Adler:2002tq}. However, the information extracted from high associated $\pt$ is limited; in order to investigate the mechanisms of partonic energy loss and the properties of the QGP medium, one needs to study associated particles at low $\pt$ where the background is large.

\subsection{Anisotropic flow correlations}

In relativistic heavy-ion collisions, the overlap parts of the colliding nuclei slow down and convert some of their kinetic energy into thermal energy. A large energy density and pressure are built up in the core of the overlap region, driving hydrodynamic expansion of the system into the surrounding vacuum. %(The spectator parts of the nuclei have passed by each other and gone.) 
In non-central collisions where the nuclei do not hit head-on, the overlap region has an oval shape. The pressure gradient along the short axis is larger than that along the long axis. This results in a more rapid expansion and larger particle azimuthal density along the short axis~\cite{Ollitrault:1992bk,Huovinen:2006jp}. This nonuniform azimuthal distribution is often characterized by Fourier series~\cite{Voloshin:1994mz},
\begin{equation}
\frac{d^2N(\eta,\phi,\pt)}{d\eta d\phi}=B_1\left(1+2\sum_{n=1}^{\infty}v_n(\eta,\pt)\cos n[\phi-\psi_n(\eta,\pt)]\right)\,,
\end{equation}
where $\psi_n$ is the symmetry plane for harmonic $n$. $B_1$ is the single particle average pseudorapidity density $\mean{d^2N/d\eta d\phi}$; the combinatoric two-particle density yields the background term in the r.h.s.~of Eq.~(\ref{eq:2p}).

In a symmetric collision system, such as \auau, the overlap region is symmetric on average. Odd harmonics ($n=3,5...$) were thus thought to be vanishing. However, individual nuclei are not a smooth distribution of matter, but composed of lumpy nucleons. Moreover, the interaction strength between two nucleons is not fixed but fluctuating quantum mechanically. (Classical physics language is used here even though it may not be correct at low $\pt$, where the nuclei interact coherently rather than through multiple nucleon-nucleon interactions. However, the lumpiness of nuclei may still have an effect.) On a collision-by-collision basis, there is no symmetry for the overlap region. All orders of harmonics ought to exist. The $v_1$ component was realized long ago~\cite{Reisdorf:1997fx}, even in the paradigm of smooth nuclei, because at any given rapidity off mid-rapidity, the amounts of ``stopped'' matter from the two nuclei are asymmetric. Only at mid-rapidity, $v_1$ is zero for smooth nuclei. In fact, for the exactly same reason, other odd harmonics (such as $v_3$) should have been realized to be non-zero as well, even with smooth nuclei~\cite{Xiao:2012uw}. %Unfortunately this was not the case.

Effect of fluctuations in the collision geometry was pointed out long ago~\cite{Bhalerao:2006tp,Andrade:2006yh,Alver:2008zza}. It was thought that the effect on anisotropic flow was small and has been neglected. It was only recently demonstrated by transport model simulations~\cite{Alver:2010gr,Alver:2010dn,Petersen:2010cw,Xu:2011fe}, followed by ``modern'' event-by-event hydrodynamic calculations~\cite{Holopainen:2010gz,Qin:2010pf,Schenke:2010rr,Qiu:2011iv,Schenke:2011bn,Qiu:2011hf,Schenke:2012wb}, that the odd harmonics can be large. %In ultra-central collisions, they can be as large as, or even larger than elliptic flow~\cite{Ma:2010dv,CMS:2013bza}. 
It was explicitly demonstrated~\cite{Takahashi:2009na,Hama:2009vu,Werner:2010aa,Andrade:2010xy,Qian:2012qn} that hydrodynamic evolutions with hot flux tubes can generate odd harmonic anisotropies. 

The harmonic symmetry planes, $\psi_n$, are not measured. They are experimentally estimated by particle distributions in the final state, exploiting the very fact that particle distributions are anisotropic--the direction of largest particle emission for a given harmonic $n$ is $\psi_n$. Due to finite particle statistics event-by-event, this estimate is not precisely $\psi_n$ but has a resolution; this can be steadily corrected~\cite{Poskanzer:1998yz}. A more difficult issue, however, is the contamination of particle correlations unrelated to the global event-wise flow correlations~\cite{Borghini:2000cm,Wang:2008gp}. Those few-body correlations are often referred to as nonflow, in contrast to flow. This contamination arises because the test particle used in flow analysis is correlated with one or a few of the other particles that are included in the event plane reconstruction.

Anisotropies are now often measured~\cite{Chatrchyan:2011eka,ALICE:2011ab,Adare:2011tg,Aamodt:2011by,ATLAS:2012at,Chatrchyan:2012wg,Adamczyk:2013waa} by two-particle correlation method, instead of the event-plane method, via the l.h.s.~of Eq.~(\ref{eq:2p}). These two methods are approximately equal; both use particle correlations. In the two-particle correlation method, particles are correlated pair-wise and the ``single-particle'' anisotropies are extracted from the Fourier coefficients. Nonflow contamination comes from the fact that some of those pairs are correlated due to physics other than hydrodynamic flow, the second term in the r.h.s.~of Eq.~(\ref{eq:2p}). The extracted anisotropies are thus not precisely the intended event-wise flow anisotropy of $V_n$, the first term in the r.h.s.~of Eq.~(\ref{eq:2p}).

Because flow is a many-particle correlation and nonflow is a few-particle correlation, one can effectively reduce nonflow contaminations by using multi-particle correlations~\cite{Borghini:2001vi,Borghini:2000sa}, such as 4-particle cumulant method. However, due to flow fluctuations, such method does not help deduce the flow background for two-particle correlations--two-particle cumulant is the mean of square, $\mean{v_n}^2+\sigma_{v_n}^2$, which is the background to jet-like correlations; but flow fluctuations affect the 4-particle cumulant such that it is smaller than the average $\mean{v_n}^4$~\cite{Borghini:2001vi}.

Jet-like correlations and two-particle flow background are, thus, effectively measured by the ``same'' two-particle correlation method. If a jet-like trigger-associated pair correlation is measured, and the ``flow'' background is also measured by the same trigger-associated pairs, then there would of course be no signal correlation (nonflow or jet-correlation signal) left. Experimentally, the trigger and associated particle anisotropies are measured by two-particle correlations with other particles (called reference particles). After background subtraction, the trigger-associated jet-like correlation signal is effectively the trigger-associated pair nonflow minus those of the trigger-reference and associated-reference pairs~\cite{Wang:2013qca}.

How does one reduce nonflow contamination in anisotropy measurements so that the measured anisotropies reflect truthfully the hydrodynamic flows? Nonflow correlations are typically short-ranged, for example, nonflow correlations from boosted resonance decays, local charge conservation~\cite{Bozek:2012en}, intra-jet correlations (jet of particles collimated in angle), and Henry-Brown-Twiss quantum interference~\cite{Lisa:2005dd} are all short-ranged in angle. Thus, a pseudorapidity gap is often applied to measure azimuthal anisotropies. There are, however, long-range nonflow correlations, such as inter-jet correlations (dijet are not strongly correlated in $\eta$ due to the random sampling of the underlying parton longitudinal kinematics) and momentum conservation (the total longitudinal momentum of measured particles is balanced by the unmeasured large momenta of particles close to the beam axis). These long-range nonflow correlations are likely smaller than the short-range nonflow correlations that are removed by a large $\deta$-gap~\cite{Xu:2012ue}.

How does one assess the magnitude of remaining nonflow contaminations? One proposed way was to study the deviation of two-particle correlation Fourier coefficients from the product of two single-particle $v_n$'s--factorization test. Experimental measurements indicate that the two-particle Fourier coefficients are factorized to a high degree at low to intermediate $\pt$~\cite{ALICE:2011ab}. This indicates that nonflow contaminations may be small. However, it was pointed out that pure jet-correlations from Pythia simulation of \pp\ collisions can yield correlations that are approximately factorized~\cite{Kikola:2011tu}, questioning factorization as a sufficient way to gauge nonflow contaminations. Moreover, it was shown that even pure hydrodynamic two-particle cumulant flow may not be factorisable~\cite{Gardim:2012im}. The harmonic symmetry planes can be decorrelated over $\eta$~\cite{Bozek:2010vz,Petersen:2011fp,Xiao:2012uw}; the event planes do not have to be the same at different $\eta$'s. This is understandable, perhaps should be expected, if geometry fluctuations play an important role in generating final-state particle anisotropy. Even with smooth geometry (no fluctuations) the $v_3$ event planes can be out of phase between forward and backward rapidities~\cite{Xiao:2012uw}. Furthermore, event-by-event hydrodynamic calculations suggest that symmetry planes depend on particle $\pt$ even at the same $\eta$~\cite{Heinz:2013bua}. This is likely a pure fluctuation effect--the response to the fluctuating geometries may differ for different physics mechanisms, responsible for particle production at different $\pt$'s.

One obvious question is, then, why factorization is so good when event-planes are badly decorrelated over $\eta$ or $\pt$? One possible reason may be that the event-plane correlations could be approximately factorized~\cite{Wang:2013qca}:
\begin{equation}
\mean{\cos n(\psi_n^{(1)}-\psi_n^{(2)})}\approx\mean{\cos n(\psi_n^{(2)}-\psi_n^{(3)})}\mean{\cos n(\psi_n^{(3)}-\psi_n^{(1)})}\,,
\label{eq:EPfact}
\end{equation}
where the superscripts indicate the phase-space regions where the event planes are determined. The event-plane decorrelation may revive the question: what is the measured anisotropy? Should anisotropic flow be measured by two-particle correlations within the same $\eta$ and $\pt$ bin? Or should it be measured with $\eta$-gap and different $\pt$ bins? There may be no answer to the question what the real flow is, but it is clear that comparisons between data and hydrodynamic calculations should be done within the same phase space with the same analysis method~\cite{Adamczyk:2013waa}. On the other hand, the effect of flow angle decorrelations  may not be as severe on flow background subtraction in jet-like correlations~\cite{Wang:2013qca}. The subtracted background should be the measured two-particle anisotropy including the event-plane decorrelation effect. However, this may require that the same $\deta$-gap be applied in the jet-like correlation and anisotropic flow measurements. The decorrelation over $\pt$ may be more serious because in flow background measurement a reference particle $\pt$ range is often chosen away from the that of the test particle~\cite{Poskanzer:1998yz,Wang:2013qca}--the breakdown of event-plane factorization of Eq.~(\ref{eq:EPfact}) could introduce systematics.

It should be noted that the measured $v_n$ are not necessarily of hydrodynamic origin. The large energy density regions (hot spots) can be a result of (mini-)jet energy deposition in the medium~\cite{Ma:2010dv,Werner:2012xh}. Their evolutions within hydrodynamics contribute to the generation of anisotropic harmonics in the final state. Jet-induced nonflow correlations and hydrodynamic flow correlations can become so entangled that it may be nearly impossible to tell them apart. 
%With all harmonic $v_n$ being possible and the two-particle correlation method bing used for both $v_n$ and jet-like correlation measurements, separation between hydrodynamic flow and jet-like correlations may be impossible at low-intermediate $\pt$. PHOBOS has studied nonflow correlations using a cluster model~\cite{Alver:2010rt}. 
Several authors have investigated data-driven methods to separate flow and nonflow correlations~\cite{Xu:2012ue,Yi:2012gx,Jia:2013tja}. A clean separation of nonflow from flow is far from completion.
%Nonflow may be small at low-intermediate $\pt$ so nonflow contamination on flow measurement may be small. Jet-like correlation analysis is intrinsically looking for nonflow effect; non-flow contaminations in background subtraction could introduce large systematic effects.

\subsection{Jet-like correlations}

Jet-like correlations have been extensively measured at RHIC for charged hadrons~\cite{Wang:2013qca}. Most of the published work were prior to the realization of non-vanishing odd-harmonic anisotropic flows. Only elliptic flow was subtracted. Dihadron correlations at high trigger and associated $\pt$ were found to be unmodified on the near side (consistent with surface emission of measured high-$\pt$ particles) and strongly suppressed on the away side (due to jet-quenching)~\cite{Adler:2002tq,Adams:2006yt}. At low and intermediate $\pt$, novel structures were observed after elliptic flow subtraction~\cite{Wang:2013qca}. Namely, a long-range pseudorapidity correlation on the near side of a high-$\pt$ trigger particle~\cite{Wang:2004kfa,Adams:2005ph,Abelev:2009af,Wang:2008zzh,Alver:2009id,Abelev:2009jv}, and a double-peak structure at $\dphi-\pi\approx\pm\pi/3$ on the away side~\cite{Wang:2004kfa,Adams:2005ph,Adler:2005ee,Adare:2006nr,Adare:2008ae,Aggarwal:2010rf,Abelev:2008ac}. While the effect of $v_3$ on correlations of high-$\pt$ associated particles is negligible due to the small background level, the effect of $v_3$ cannot be neglected at low-intermediate $\pt$. The main features of the observed correlation structures are consistent with a sizable background from $v_3$. Quantitatively, how much jet-like correlations survive after subtraction of $v_3$ and other non-negligible harmonics remains an open question. This further work of $v_3$ subtraction should be carried out in the future. It will require careful measurements of anisotropic flow harmonics and assessment of nonflow contaminations. Given unambiguous experimental evidence of jet quenching at high $\pt$~\cite{Adler:2002tq,Adams:2006yt,Adare:2010ry}, effects of jet-medium interactions should remain in jet-like correlations and should potentially provide indispensable information to further understand QCD at extreme conditions.

\begin{figure*}
\begin{center}
\includegraphics[width=1.04\textwidth]{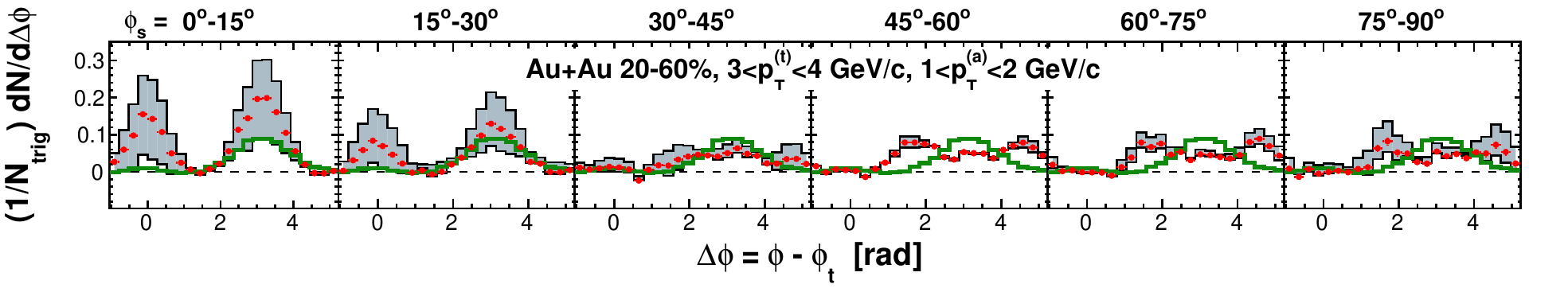}
\caption{Dihadron correlations at $|\Delta\eta|>0.7$ in 20-60\% \auau\ collisions at 200~GeV with trigger particles in six slices of azimuthal angle relative to the event plane, $\phi_s=|\phi_t-\psi_{EP}|$. The trigger and associated $\pt$ ranges are $3<\ptt<4$~\gev\ and $1<\pta<2$~\gev, respectively. The $v_2$, $v_3$, and $v_4$ backgrounds have been subtracted. The histograms with shaded area inbetween used, respectively, two- and four-particle cumulant $v_2$ in background subtraction.
%Systematic uncertainties are shown in the black histograms due to flow subtraction and in the horizontal shaded band around zero due to ZYAM background normalization. 
%Systematic uncertainties due to flow subtraction are shown in the shaded areas. 
The inclusive dihadron correlations from minimum-bias \dau\ collisions (thick green histograms) are superimposed for comparison. From STAR~\cite{Agakishiev:2010ur,Wang:2013jro}.}
\label{fig:EPdep}
\end{center}
\end{figure*}

The additional $v_3$ background subtraction has been carried out in the dihadron correlation analysis with respect to the event plane by STAR and PHENIX~\cite{Agakishiev:2010ur,PHENIX:2013kia,Wang:2013jro}. The two-particle cumulant $v_n$, with $\eta$-gap corresponding to that used in the jet-like correlation analysis, are subtracted. These should be the largest possible anisotropic flow to be subtracted. The background subtracted jet-like correlation signal by STAR~\cite{Agakishiev:2010ur,Wang:2013jro} is shown in Fig.~\ref{fig:EPdep} for trigger particle from in-plane to out-of-plane direction. The systematic uncertainties correspond to subtraction of elliptic flow from the two- and four-particle cumulant methods. Interesting correlation structures remain; while the near-side large-$\deta$ ridge (peak at $\dphi\sim0$) is mostly gone, the out-of-plane away-side correlation is broad and perhaps double-peaked. %This could be a manifestation of jet modification by medium
It is probably too early to tell whether the remaining structure is a manifestation of jet modification by medium, 
%an over- or under-subtraction of nonflow contaminated anisotropic flow backgrounds, 
or systematics unidentified so far. One limitation of the STAR study is the limited $\eta$ acceptance where effects of jets on the determination of the event plane are not fully quantified. %This cannot be satifactorily answered with the present STAR detector, but the 
The wide $\eta$ coverage of LHC experiments should have advantage in such measurements. 

The LHC experiments have so far exploited the two-particle correlation mainly as a way to measure anisotropic flow~\cite{Chatrchyan:2011eka,ALICE:2011ab,Aamodt:2011by,ATLAS:2012at,Chatrchyan:2012wg}. The nonflow contaminations, such as that from away-side inter-jet correlations at large $\deta$, have not been fully investigated. The full potential of the LHC data has not yet been realized; efforts should be put into measurements of anisotropic flow as accurately as possible and subtraction of flow to extract jet-like correlation signals.

One measurement that is relatively insensitive to flow background subtraction is near-side intra-jet correlations by the method of subtracting large-$\deta$ correlations from small-$\deta$ correlations. This is because flow is rather constant over the relevant $\eta$ ranges and thus canceled in such subtraction. Measurements at RHIC show that the near-side correlations are rather invariant over collision systems and centralities~\cite{Abelev:2009af,Agakishiev:2010ur}. This is consistent with surface emission and in-vacuum fragmentation of high-$\pt$ particles. However, STAR has lowered the trigger and associated $\pt$ to as low as 1.5 \gev. It is found that the correlations are still the same in \auau\ and \dau\ collisions~\cite{Konzer:QM2012}. It is surprising given that a relatively large fraction of $\pt\sim1.5$~\gev\ particles are part of the bulk medium; not all of them are jet related, and they should have emerged through the interior and have interacted strongly in the medium. The data appear to suggest that those final state interactions have no effect on their angular correlations. 

\section{The ridge in small systems\label{sec:ridge}}

Small-system \pp, \pA, and \dA\ collisions are usually used as reference to heavy-ion collisions. Deviations of heavy-ion data from the properly normalized \pp\ and $p(d)A$ data--for example, the nuclear modification factor $R_{\Aa}$--are taken as evidence of final-state interactions~\cite{Jacobs:2004qv}. Rather surprisingly, a long-range pseudorapidity ridge correlation, very much like that in heavy-ion collisions, was observed in high-multiplicity \pp\ collisions by the CMS experiment at the LHC~\cite{Khachatryan:2010gv,Li:2012hc}. Subsequently, a similar ridge, much stronger in magnitude, was observed in \ppb\ collisions by CMS, ALICE, and ATLAS~\cite{CMS:2012qk,Abelev:2012ola,Aad:2012gla}. It should be noted that the ridge referred to in heavy-ion collisions was measured after the subtraction of elliptic flow background, whereas the LHC \pp\ and \ppb\ ridge was measured with a uniform background subtraction. As aforementioned, the heavy-ion ridge is primarily attributed to triangular flow $v_3$. It is, therefore, attempting to attribute the \pp\ and \ppb\ ridge to elliptic (and triangular) flow, given the similarity of the observations.

\begin{figure}[hbt]
\begin{center}
\includegraphics[width=0.4\textwidth]{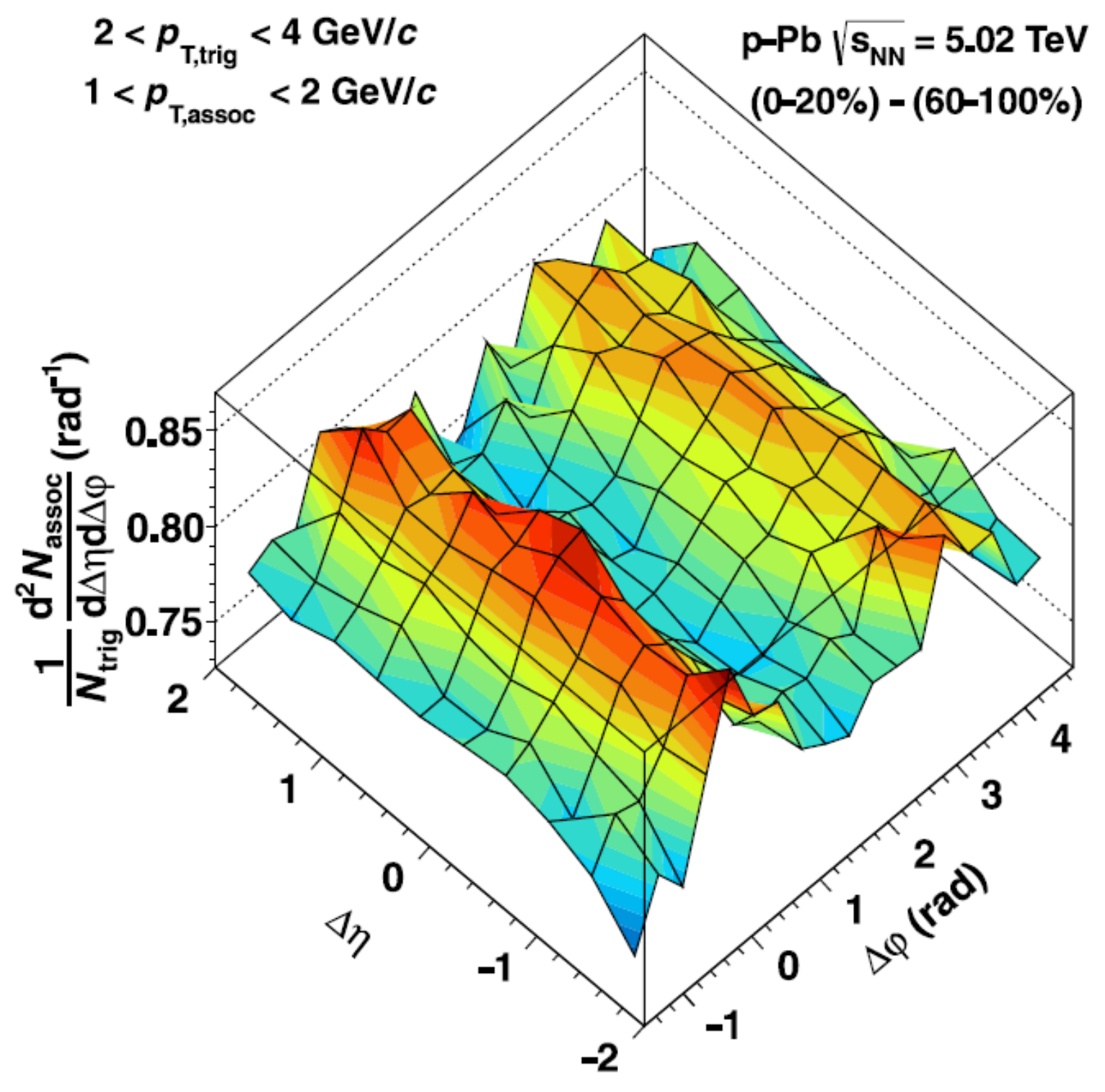}
\includegraphics[width=0.42\textwidth]{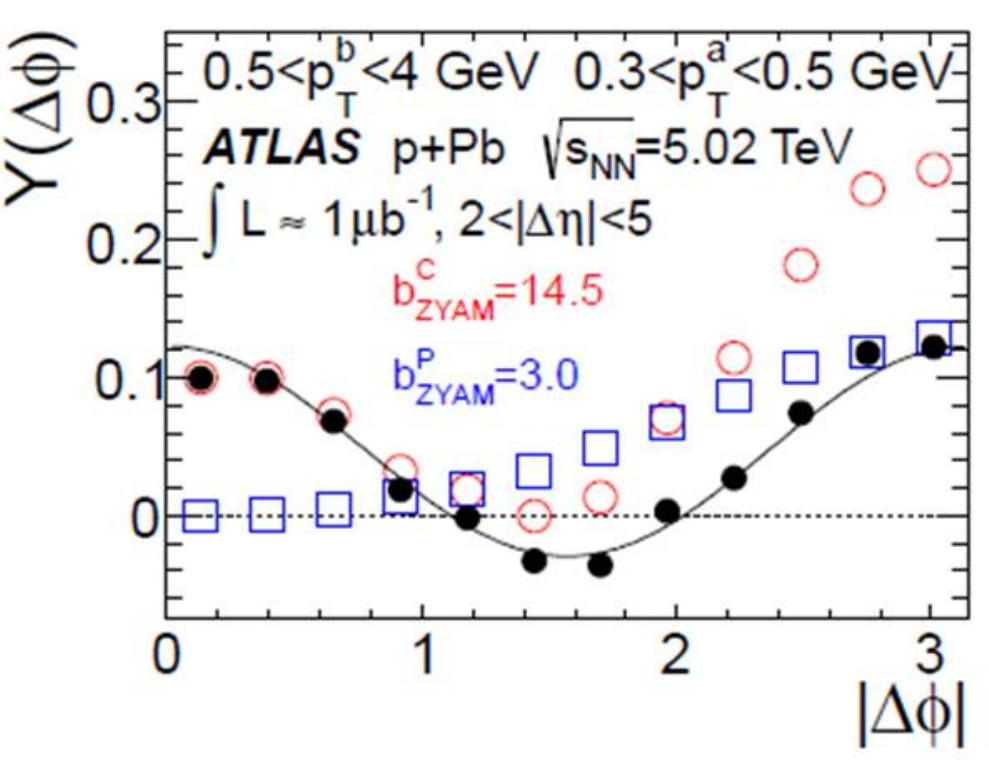}
\end{center}
\caption{(Left panel) ALICE result~\cite{Abelev:2012ola} on the difference of dihadron correlations between central 0-20\% and peripheral 60-100\% \ppb\ collisions. The trigger and associated $\pt$ ranges are $2<\ptt<4$~\gev\ and $1<\pta<2$~\gev, respectively. (Right panel) ATLAS result~\cite{Aad:2012gla} of dihadron correlations in peripheral (48-100\%) and central (0-2\%) \ppb\ collisions, and their difference; $0.5<\ptt<4$~\gev\ and $0.3<\pta<0.5$~\gev.
%(Right panel) PHENIX result~\cite{Adare:2013piz} of dihadron correlations in peripheral (50-88\%) and central (0-5\%) \dau\ collisions and their difference; $0.5<\ptt<0.75$ and $0.75<\pta<1$~\gev.
}
\label{fig:sub}
\end{figure}

While the large $\deta$ near-side correlation is unlikely due to jets, the away-side correlations in \pp\ and \pA\ collisions are dominated by inter-jets. If jet correlations are the same in low- and high-multiplicity events, then additional correlation structures can be easily identified after subtracting the two-particle azimuthal correlation in peripheral collisions from that in central collisions. Such a subtraction by ALICE and ATLAS revealed a back-to-back double ridge~\cite{Abelev:2012ola,Aad:2012gla}. This is shown in Fig.~\ref{fig:sub}. The double ridge is of an approximate $\cos2\dphi$ shape, reminiscent of an elliptic flow contribution. In realty, selecting events according to final-state multiplicity can bias jet correlations in those events. This is because, in those ``small'' events, the contribution of jet-related particles can be important to the overall multiplicity, and because large multiplicity events are more likely associated with large energy dijets. These biases are largely reduced by using multiplicity measurement far removed in $\eta$ from the jet-correlation measurement. There might still be biases of about 10-20\% remaining in the measured double ridge at the LHC~\cite{Aad:2012gla,Abelev:2012ola}.

PHENIX has followed the same subtraction procedure to analyze their \dau\ data~\cite{Adare:2013piz}. It was found that the ``central minus peripheral'' difference exhibits the characteristic $\cos2\dphi$ modulation in their limited $\deta$ acceptance of 0.48-0.7. It is unclear whether the signal is due to a multiplicity bias of jet-correlations or evidence of new physics. In fact, a difference in azimuthal correlations between \pp\ and \dau\ collisions was previously measured by STAR using a cumulant variable~\cite{Adams:2004bi}. %It was attributed to different nonflow (e.g.~jet) correlations between the two collision systems. 
%
%PHENIX has estimated multiplicity bias to jet-correlations, and found it is only an 10\% effect~\cite{Adare:2013piz}. 
Multiplicity biases to jet correlations can be estimated from $\deta$ correlations by STAR with a relatively large $\deta$ acceptance. Preliminary STAR data~\cite{Wang:IS2013} suggest that most of the signal in ``central minus peripheral'' correlations at mid-rapidity in the main Time Projection Chamber (TPC) can be explained by centrality selection biases, e.g.~by multiplicities measured in the Au-beam direction forward TPC (FTPC).

\begin{figure*}[ht]
\begin{center}
\includegraphics[width=0.45\textwidth]{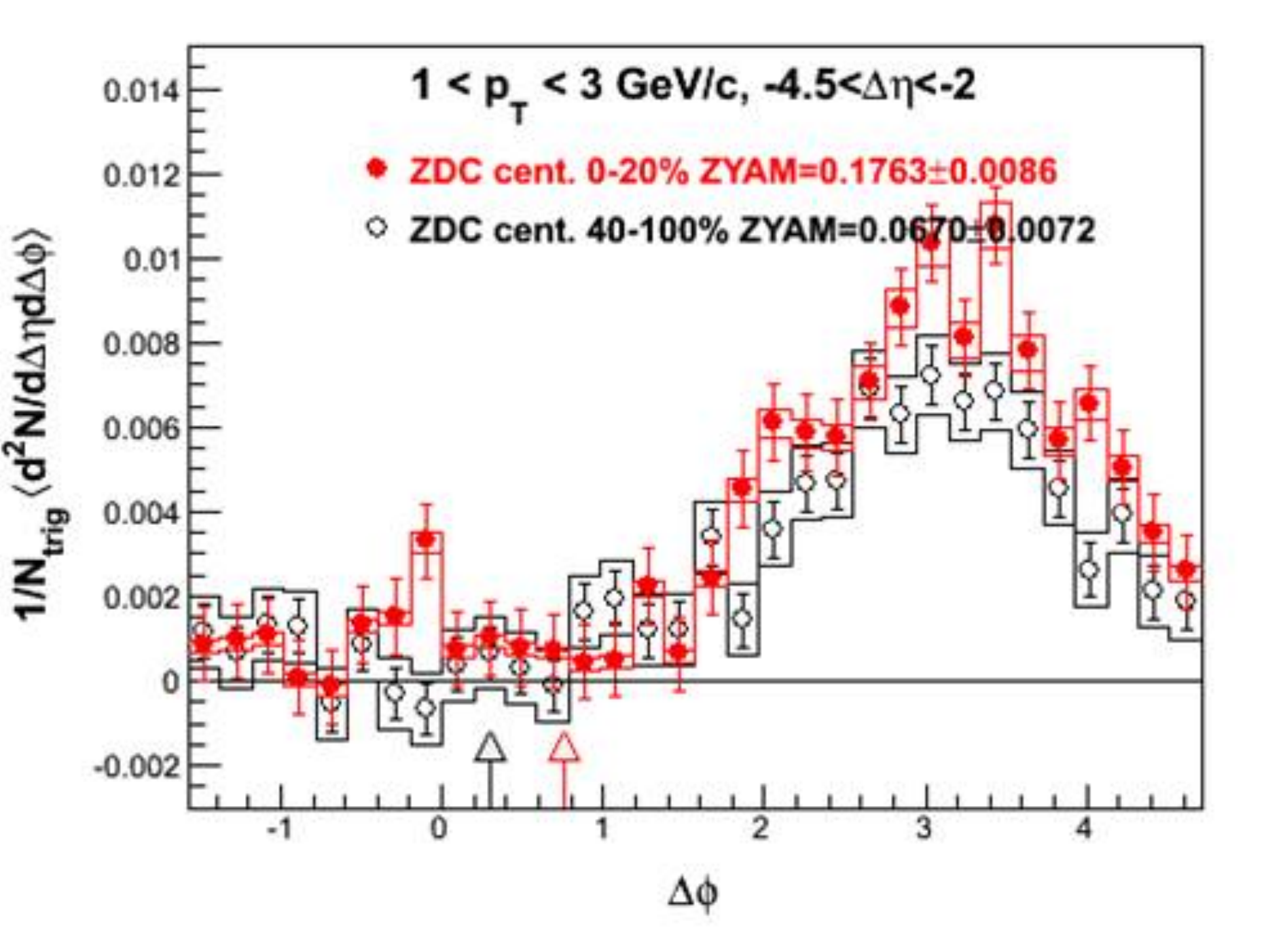}
\includegraphics[width=0.40\textwidth]{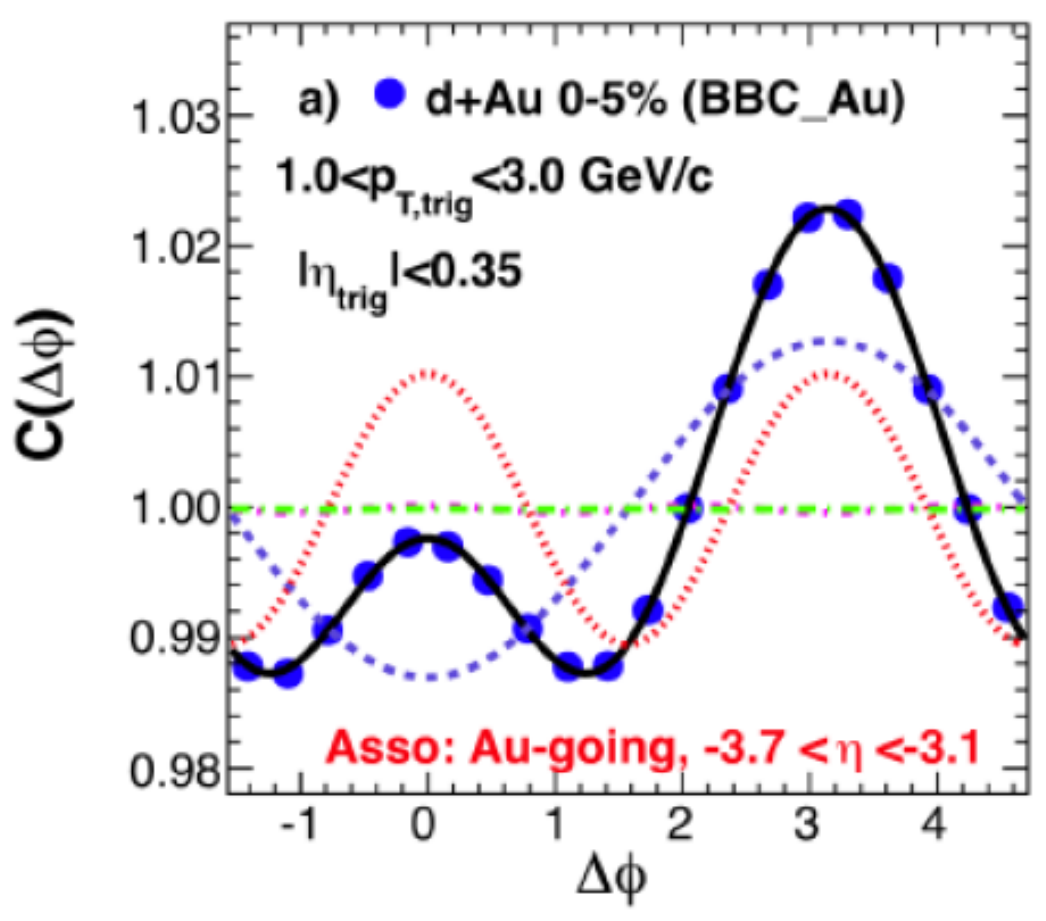}
\caption{(Left panel) Dihadron correlations between TPC trigger ($1<\ptt<3$~\gev, $|\eta|<1$) and FTPC associated particles ($1<\pta<3$~\gev, $-3.8<\eta<-2.8$) from STAR~\cite{Wang:IS2013}. Both central and peripheral \dau\ data are shown; centrality is determined by measurement in the Au-going Zero Degree Calorimeter (ZDC). Error bars are statistical and the histograms show systematic uncertainties. (Right panel) Correlation function between mid-rapidity trigger ($1<\ptt<3$~\gev, $|\eta|<0.35$) and MPC energy ($-3.7<\eta<-3.1$) from PHENIX~\cite{Huang:HP2013}. Top 5\% \dau\ data are shown; centrality is determined by measurement in the Au-going Beam-Beam Counter (BBC). Statistical errors are smaller than the point size.}
\label{fig:forward}
\end{center}
\end{figure*}

STAR has measured TPC-FTPC correlations with an $\deta$ gap of approximately 3. There appears to be a non-zero near-side correlation signal at this large $\deta$~\cite{Wang:IS2013} (see left panel of Fig.~\ref{fig:forward}). PHENIX has measured a prominent near-side correlation peak at large $\deta$ using their backward Muon Piston Calorimeter (MPC) with trigger particles at mid-rapidity~\cite{Huang:HP2013} (see right panel of Fig.~\ref{fig:forward}). The PHENIX MPC detector measures energy flux from particles of all $\pt$, whereas the STAR FPTC detector measures particle tracks with momentum information. At this large $\deta$, correlated particles are unlikely from the triggered near-side jet. The underlying physics mechanism for the large $\deta$ near-side correlations, however, remains an open question. Measurement of the composition of those correlated particles may be illuminating. It is worth to note that there are real physics differences in \dau\ collisions between forward ($d$) and backward (Au) directions. An enhanced away-side correlation is measured in backward rapidities in high-multiplicity relative to low-multiplicity \dau\ collisions (see left panel of Fig.~\ref{fig:forward}), while a depletion is measured in the forward direction~\cite{Braidot:2010ig,Wang:IS2013}. It is unclear whether these away-side differences are due to multiplicity biases, initial-state multiple scattering, or differences in the parton distributions in $d$ and Au.

The LHC experiments have extracted anisotropy parameters from the double-ridge azimuthal correlation functions~\cite{Aad:2012gla,Abelev:2012ola,Aad:2013fja,Chatrchyan:2013nka}. The extracted $v_2$ is rather weakly dependent of the measured mid-rapidity multiplicity (an indicator of collision violence). Even the four-particle cumulant results are non-zero and rather insensitive to multiplicity. From the \ppb\ data, a finite non-zero two-particle $v_3$ is also observed. Surprisingly, the two-particle $v_3$ from \ppb\ and peripheral \pbpb\ collisions follow an identical curve as a function of multiplicity. This is naively unexpected given the presumably very different fluctuating geometries in \ppb\ and \pbpb\ collisions.

%\begin{figure*}[ht]
%\begin{center}
%\includegraphics[width=0.4\textwidth]{CMS_pPb_v2.pdf}
%\includegraphics[width=0.39\textwidth,height=0.335\textwidth]{CMS_pPb_v3.pdf}
%\caption{(Left panel) The two- and four-particle cumulant measurements of $v_2$ in \ppb\ collisions by ATLAS~\cite{Aad:2013fja} and CMS~\cite{Chatrchyan:2013nka}. (Right panel) The two-particle cumulant measurement of $v_3$ in \ppb\ and \pbpb\ collisions as a function of event multiplicity by CMS~\cite{Chatrchyan:2013nka}. Figure from Ref.~\cite{Chatrchyan:2013nka}.}
%\label{fig:CMSvn}
%\end{center}
%\end{figure*} 

ALICE has measured the double-ridge correlation with identified associated pions, Kaons, and protons~\cite{ABELEV:2013wsa}. The second harmonic Fourier coefficient as a function of $\pt$ exhibits patterns similar to hydrodynamic flow. This is a strong indication that hydrodynamic flow may be indeed the underlying physics mechanism for the observed long-range ridge correlations at the LHC.

In fact, hydrodynamic calculations of \pp\ and \ppb\ collisions can describe the main features of the measured ridge correlations~\cite{Bozek:2010pb,Werner:2010ss,Bozek:2011if,Bozek:2012gr}. The extracted anisotropy parameters can be semi-quantitatively reproduced~\cite{Adare:2013piz}. On the other hand, if an anisotropic flow is attained by high-$\pt$ particles, nearly as large as that in heavy-ion collisions, then there should be significant effect of jet quenching in \ppb\ collisions. However, no evidence of strong jet quenching has been observed in \ppb\ collisions~\cite{ALICE:2012mj}. 

Hydrodynamic flow is not the only possible explanation for the ridge in small systems. Another possible explanation is the color glass condensate (CGC)~\cite{JalilianMarian:2005jf,Gelis:2010nm}. The gluon density at small Bjorken $x$ is saturated below a certain momentum, called the saturation scale~\cite{McLerran:1993ni,McLerran:1993ka,Iancu:2001md,Gelis:2007kn,Gelis:2010nm}. In the CGC framework, the two-gluon density is enhanced at small relative azimuthal angle, and it is shown that the CGC effective field theory can reproduce the \pp\ ridge~\cite{Dumitru:2010iy,Dusling:2012iga}. The same framework can also reproduce the back-to-back double ridge in \ppb\ collisions~\cite{Dusling:2012cg,Dusling:2012wy,Dusling:2013oia} by varying the saturation scales in proton and lead where the latter depends on the number of participant nucleons from the lead nucleus. However, CGC may not naturally describe the third harmonic Fourier coefficient observed in experimental data. 

\section{Summary and future prospects\label{sec:future}}

This article reviews the current status of jet-quenching and correlations in relativistic heavy-ion collisions. It covers three areas of extensive recent research: high-$\pt$ hadron and jet suppression as well as jet and direct photon induced correlations, interplay between jet-like correlations at low to intermediate $\pt$ and collective anisotropic flow, and long-range pseudorapidity ridge correlations in small systems of \pp\ and \pdA\ collisions.

The evidence of high-$\pt$ jet-quenching is unambiguous. Jet reconstruction is limited by heavy-ion background and fluctuations, the subtraction of which is not discussed in detail in this review. The biases due to these limitations in the reconstructed jets are exploited to study jet-quenching. It would be valuable to reconstruct jets as unbiased as possible in the study of jet-quenching and partonic energy loss mechanisms. This is clearly one of the future directions in jet reconstruction.

The mechanisms of partonic energy loss and the role of medium response have been elusive. Measurements of how the lost energy is redistributed suffer from large uncertainties in the underlying flow background subtraction. How jets interact with the QGP medium and how the medium responds to jet energy loss are important questions central to the field of study of QCD matter under extreme conditions. Future work must address the question of nonflow contaminations in flow measurements to a high precision. This will allow a more precise subtraction of flow background at low-intermediate $\pt$ in jet-like correlation measurements.

Direct photons are immune from anisotropic flow background. Future high statistics data on direct $\gamma$-hadron and $\gamma$-jet correlations may be essential to address the question of jet-medium interactions and the properties of the hot dense QCD medium.

The observation of the ridge in small systems is unexpected. The similarity to the ridge observed in heavy-ion collisions is surprising, where the ridge is primarily attributed to triangular flow. Is Nature kind to simply tell us that the ridge in small systems is due to hydrodynamic flow? Does hydrodynamics make sense in \pp\ and \pdA\ collisions? Is it possible to have large high-$\pt$ anisotropy and little jet-quenching? CGC seems to be able to explain the second Fourier harmonic component of the ridge correlations in small systems. Can CGC also explain the third Fourier component observed in \ppb\ collisions? Can CGC together with gluon fragmentation account for the identified particle ridge correlations? Future theoretical work on hydrodynamics and CGC should explore the parameter space to make predictions with theoretical systematic uncertainties. Future experimental work should focus on more extensive studies of the ridge phenomena in small systems to arrive at the most natural explanation--hydrodynamics, CGC, or possibly other physics mechanisms.

\acknowledgments
I thank my colleagues for discussions and collaboration. This work is supported by US Department of Energy Grant No. DE-FG02-88ER40412.

\bibliographystyle{pramana}
\bibliography{review}
\end{document}